# Observations of Ultrafast Superfluorescent Beatings in a Cesium Atomic Vapor Excited by Femtosecond Laser Pulses

Gombojav O. Ariunbold,[1,*] Vladimir A. Sautenkov,[2] Hebin Li,[3] Robert K. Murawski,[4] Xi Wang,[5] Miaochan Zhi,[6] Tuguldur Begzjav,[7] Alexei V. Sokolov,[8] Marlan O. Scully,[8] Yuri V. Rostovtsev[9]

[1]Department of Physics and Astronomy, Mississippi State University, Mississippi State, MS 39762, USA
[2]Joint Institute for High Temperatures of Russian Academy of Sciences, Moscow 125412, Russia
[3]Department of Physics, Florida International University, FL 33199, USA
[4]Department of Physics, Drew University, Madison, NJ 07940, USA
[5]Global Oncology One Inc., Houston, TX 77054, USA
[6]U.S. Food and Drug Administration, Silver Spring MD 20993, USA
[7]Department of Physics, National University of Mongolia, Ulaanbaatar 14200, Mongolia
[8]Institute for Quantum Studies and Department of Physics, Texas A & M University, College Station TX 77843, USA
[9]Department of Physics, University of North Texas, Denton TX 76203, USA
*Corresponding author: g.o.ariunbold@gmail.com

**Spontaneous emission from individual atoms in vapor lasts nanoseconds, if not microseconds, and beatings in this emission involve only directly excited energy sublevels. In contrast, the superfluorescent emissions burst on a much-reduced timescale and their beatings involve both directly and indirectly excited energy sublevels. In this work, picosecond and femtosecond superfluorescent beatings are observed from a dense cesium atomic vapor. Cesium atoms are excited by 60-femtosecond long, 800 nm laser pulses via two-photon processes into their coherent superpositions of the ground 6S and excited 8S states. As a part of the transient four wave mixing process, the yoked superfluorescent blue light at lower transitions of 6S – 7P are recorded and studied. Delayed buildup time of this blue light is measured as a function of the input laser beam power using a high-resolution 2 ps streak camera. The power dependent buildup delay time is consistently doubled as the vapor temperature is lowered to cut the number of atoms by half. At low power and density, a beating with a period of 100 picoseconds representing the ground state splitting is observed. The autocorrelation measurements of the generated blue light exhibit a beating with a quasi-period of 230 fs corresponding to the splitting of the 7P level primarily at lower input laser power. Understanding and, eventually, controlling the intriguing nature of superfluorescent beatings may permit a rapid quantum operation free from the rather slow spontaneous emission processes from atoms and molecules.**

## 1. INTRODUCTION

Spontaneous emission from excited atoms is a relatively slow and hardly manipulative process. Spontaneous emission is scaled in nanoseconds or microseconds intrinsically depending on the bandwidth of the excited energy levels of atoms and molecules in gas phase. With a laser excitation, the atoms and molecules are prepared in their collective state and eventually emit synchronously [1-6]. Synchronized spontaneous emission is commonly called superfluorescence (SF) and that occurs in a much-reduced timescale compared to that for unsynchronized spontaneous emission [7,8]. SF is observed in various atomic and molecular vapors [9-14] as well as other exotic materials [15-20]. A two-level model predicts the peak intensity and delay of SF pulses that are proportional to $N^2$ and $N^{-1}$, respectively, where N is the number of two-level atoms in the ensemble [7,8]. A three-level model exhibits the so-called cascade SF [21,22]. This process can be understood as a transient four-wave mixing process that involves two input laser pulses and two emitted pulses. If these two emitted pulses in the cascade transition did occur simultaneously then the emission in the lower transition is referred to as a yoked superfluorescence (YSF) [13,23,24].

Common beatings that observed in spontaneous emissions involve directly excited energy sublevels only, however, the beatings that measured in SF emissions can involve both directly and also indirectly excited energy sublevels [25,26] when pumped by the laser light. Another important difference between them is that the SF beating frequency can be shifted depending on many parameters including the input laser power and number density which rather obscures the beating interpretation. For example, SF beating demonstrates quasi-periodic behaviors both red and blue shifts can be expected [25, 27]. In the early works [25,26], cesium (Cs) atomic vapor was excited with nanosecond (ns) pulses and the SF emissions were detected on a nanosecond time scale. Later, a picosecond (ps) pulse excitation of the Cs vapor and the detection of the YSF on hundreds of picoseconds were reported [23,24]. However, to our knowledge, the detection of the YSF and its beating properties on a shorter time scale has not been reported yet. This can be achieved, e.g., by pumping the dense Cs atoms with a femtosecond (fs) pulses and detecting the generated light signal with an ultrafast streak camera.

In this letter, we study the YSF and its beating behaviors in a dense Cs vapor pumped by 60-fs laser pulses. This research is a complementary to our previous series of works on the cooperative emissions from a rubidium [5,6,13,28,29] and sodium vapors [14]. Particularly, our focus is to observe the ultrafast beatings. Next section introduces the experimental setup and detection methods. In section 3, the observed results will be discussed. The last section is conclusion.

## 2. MATERIALS AND METHODS

A Ti:Sapphire amplified laser system (Coherent Inc.) is used to produce a 60-femtosecond (fs) long pulses at 804 nm center wavelength (half width at half maximum – HWHM of 20 nm) at one kilohertz repetition rate. The maximum input power is about 800 mW (i.e., 800 mJ/pulse). A beam diameter is approximately half centimeter of an area of 0.2 cm$^2$. In this case, e.g., 200 mW power is converted to 0.2 mJ energy per pulse, thus, the pulse energy fluence becomes 1 mJ/cm$^2$. The laser beam enters a 3-inch long cylindrical cell with Cs atoms. The cell is heated up and stabilized (with a variation of less than 2 degrees during experiment) either at 217 C or 242 C. The main reason of this particular temperature selection is that the atomic density of approximately $3\times10^{16}$ cm$^{-3}$ at 217 C is approximately doubled (approximately $6\times10^{16}$ cm$^{-3}$) at 242 C. As sketched in Fig. S1 in Supplemental Document, the atoms are efficiently excited from the ground state 6S via two-photon absorption through the intermediate level (6P$_{3/2}$ state is not shown here) into 8S state. The laser beam center wavelength line is about 9 nm off from its two-photon resonant transition on 6S – 8S. It is important to note that 6S – 7D transition is even further away (about 19 nm off from 804 nm). Since the HWHM is about 20 nm, therefore, the processes associated with 6S – 7D are substantially diminished. However, another competitive process involving 6S$_{1/2}$, 8S$_{1/2}$, 6P$_{3/2}$, 6P$_{1/2}$ may still be present, which was studied and reported in Refs. [23,24]. The emissions (761 nm, 795 nm, 852 nm and 895 nm), in this case, fall within the near infrared region and are filtered out together with the 804 nm input beam. The emissions in transition 7P – 8S are in mid infrared region and are filtered out. In addition, our streak camera and photodiode detector sensors are less or not sensitive in this region. This process is also referred to as a transient four-wave mixing as in [23,24]. Two photons through intermediate 6P$_{3/2}$ state excite atoms. The two-photon excitation eventually triggers simultaneous emissions on both 7P – 8S and 6S – 7P transitions. The emissions in transitions 6S$_{1/2}$ – 7P$_{3/2}$ at 456 nm and 6S$_{1/2}$ – 7P$_{1/2}$ at 459 nm are spectrally resolved, see Fig. S1 (C). However, an intensity ratio between the two is significant and 456 nm is mainly contributed. Both emissions represent YSF and have a common ground state [25] therefore, we expect beating in a 456 nm emission (blue light). The splitting of 7P level is 143 cm$^{-1}$, which corresponds to 233 fs. On the other hand, the splitting of the ground state is about 9 GHz which corresponds to 108 ps beating. These beatings are observed and discussed in the following section.

The generated blue light beam profiles (approximately 4 cm diameter) are shown for several different input laser powers (from 100 mW to 450 mW with an increment of 50 mW) at the temperature 242 C in Fig. S1 (D). For low power (<250 mW) the blue light divergence angle is estimated to be less than one degree, however, for high power (>250 mW) the profile diameter is more than doubled due to ring formation. For higher power, beam profiles have distinct ring and central spot structures. More rigorous theoretical and experimental studies are needed to explore the structured beam profiles and non-monotonic power dependence (see Fig. S1 (E)) as discussed in [30]. In this work, for the sake of transparency, we mainly focus on the data collected at lower power (< 250 mW at the temperature 242 C). A detailed setup layout is sketched in Fig. S2 in Supplemental Document. The amplifier output pulses (with optional the Michelson interferometer for double pulses with a variable delay) enter the cell and generate blue light. The generated blue light is spectrally filtered and then detected either by the specially designed photodiode detectors or the 2-ps resolution streak camera. A response time of the commercially available photodiode detectors (OPT101, Thorlabs Inc.) is intentionally increased to a half millisecond as to improve signal to noise ratio by the time-integrated detection scheme. A digital storage oscilloscope (Tektronix) displays and records both input and generated light. A fast streak camera (Hamamatsu, c5680) is used to record temporal characteristics of the input and generated light.

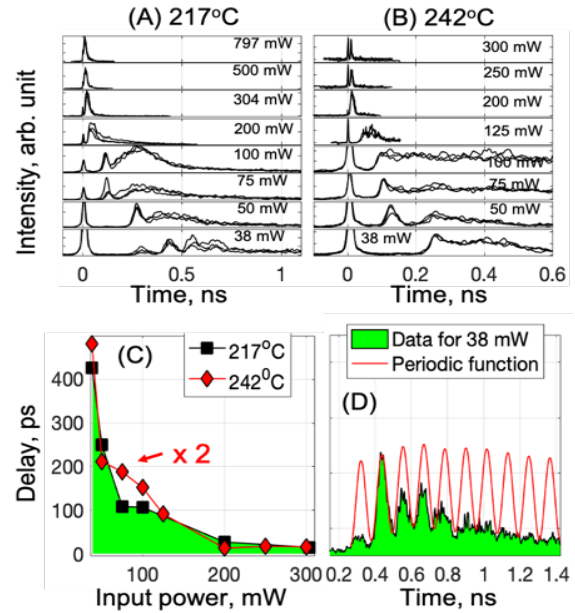

Fig. 1. Pulse temporal characteristics are recorded by the streak camera. (A) Data for 38, 50, 75, 100, 200, 304, 500 and 797 mW at temperature of 217 C. (B) Data for 38, 50, 75, 100, 200, 250 and 300 mW at temperature 242 C. (C) Scaled delay dependence on input power. Red (black) curves for temperature 217 C (242 C). (D) The selected data (black curve) for 38 mW input power at 217 C compared with a sinusoidal function (dashed curve).

## 3. RESULTS AND DISCUSSIONS

The generated blue light pulse characteristics are shown in Fig. 1. The data at 217 C (see, A) and at 242 C (see, B) and their comparison (see, C) are depicted for different input powers. The sharp peaks are input 804 nm pulses and their rising edge halves are aligned at zero. The following delayed pulses are the characteristics of the blue light pulses. As seen from both figures (A) and (B), buildup delays tend to elongate as the input power decreases. This is the main difference of the cooperative process from non-cooperative processes. For example, at temperature 242 C, the pulse shapes and delay off-sets become insensitive for relatively high input powers (within a 2 ps resolution) which indicates non-cooperative emissions, see Fig. S3. As power gradually increases above 300 mW (at 242 C), beam profiles first become non-uniform consisting of the central spot and distinct ring formations, and then for relatively high powers (>450 mW), they become nearly uniform and collimated. Next, as mentioned above, these two temperatures correspond to the density ratio of two. Thus, in Fig. 1 (B) the scaled values of positions of the pulse peak halves at rising edges as functions of variable input power are plotted. In this case, the delay values for 242 C are scaled up by a factor of two. The delay time behaviors of the two quantitatively agree which indicates that they are proportional to $N^{-1}$. Another intriguing feature of the YSF is the beating can be printed directly on the temporal pulse characteristics [25,26]. The data for 38 mW power at 217 C exhibits this feature as compared to the sinusoidal function with a period of 100 ps, see Fig. 1 (D). The splitting of the ground state $6S_{1/2}$ is 9.19 GHz [25,26] corresponding to 108 ps. It is important to note that the quasi periodicity due to the SF beatings are different from the well-known Burnham-Chiao ringing [31], which explicitly depends on square root of time as was previously observed by the authors [6, 13]. In our current experimental condition, this power of 38 mW is the lowest possible input power for the minimum observable signal intensity enough to be captured with the streak camera. All three data for this power distinctly exhibits the 100-ps beating. However, for a higher input power range, this beating, unfortunately, vanished.

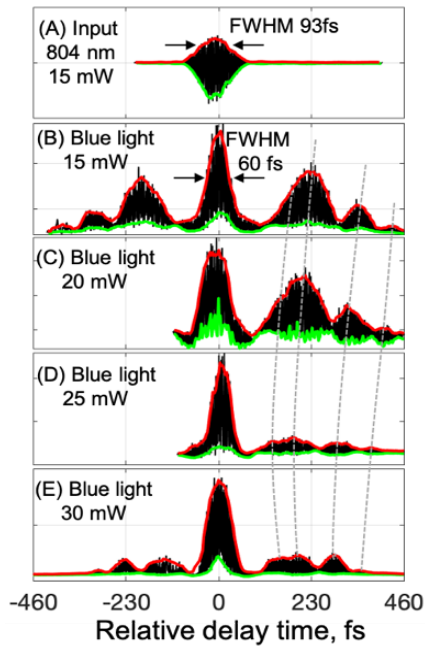

Fig.2. Interferometric autocorrelation data for input beam (A), the generated blue beam at input power of (B) 15 mW, (C) 20 mW (D) 25 mW (E) 30 mW.

The streak camera resolution used here is limited, but (interferometric) autocorrelation measurements utilizing a pair pulse with variable time delay can reveal beating further down to fs time scale [28,29]. All autocorrelation data are taken at temperature 242 C. As seen in Fig. S2, a Michelson interferometer is inserted on the path of input beam. One of the two end mirrors is facilitated to scan controlled by the AC function generator. The scope captures the signals in three channels: applied voltage, autocorrelation traces for the input and generated beams. Figure 2 displays a set of the autocorrelation data for several different input powers. In Fig. 2 (A), the autocorrelation signal for input beam is measured to be 93 fs. Assuming Gaussian pulse shapes, the pulse temporal width corresponds to 60 fs. This width is confirmed with the measurement by the commercial second order autocorrelator (APE Inc.). We emphasize here that the above interferometric autocorrelation measurement demonstrates a linear interference, which is only sensitive to spectral intensities, but not phases. This one-to-one comparison of the first-order autocorrelation to the second-order autocorrelation concludes that the spectral phase of the input laser pulse is flat. The generated blue light autocorrelations are shown in Figs. 2 (B, C, D and E) for input powers of 15, 20, 25 and 30 mW, respectively. We note that the beating feature observed here not periodic corresponding to a single narrowband beat frequency. However, in this case, an important beating information can be extracted from these data. For example, the measured autocorrelation data exhibit the repeated bumps which have also certain temporal widths. As mentioned in Ref. [27], the central bump may not be associated with SF. Interestingly, the width of the central bump is measured to be 60 fs, which is shorter than the input beam width of 93 fs. However, the observed off-center bumps are directly related to SF. Unambiguously, maxima of the second bumps are located at a relative delay about 230 fs away from the center (zero delay) position. This is expected since the splitting of 7P state is about 4.3 THz corresponding to 230 fs. Finally, we note that as input power increases the nature of SF beating is eventually washed out, see the cases (D) and (E) in Fig. 2.

## 4. CONCLUSIONS

Spontaneous emission of excited individual atoms in vapor lasts nanoseconds, if not microseconds and the beatings in it involve directly excited energy sublevels. In contrast, the superfluorescent burst occurs, e.g., on a picosecond timescale. Since the generated superfluorescent light is free from dephasing and/or spontaneous emission decay processes, the beatings in superfluorescent light can involve not only directly laser excited sublevels but also indirectly excited and even ground energy sublevels. In this work, the ultrafast superfluorescent beatings that involve both excited and ground energy sublevels are observed. Cs atoms are excited by 60-femtosecond long, 804 nm laser pulses via two-photon near resonant processes into their coherent superpositions of the ground 6S and excited 8S states. As a part of the transient four wave mixing process, the yoked superfluorescent blue light pulses at lower transitions of 6S – 7P are recorded. Delayed buildup time of this blue light is measured as a function of the input laser beam power by the use of a high-resolution streak camera. Delays are scaled appropriately with the number of atoms as varying the temperature of the vapor. At low power and density, a beating with a period of 100 picoseconds corresponding to the ground state splitting is observed. The linear autocorrelation measurements exhibit a beating with a quasi-period of 230 fs corresponding to the splitting of the 7P level primarily in the lower input laser power range. At increased laser power, the nature of the both observed beating is worn away. Temporal coherent control [32-34] using the ultrashort superfluorescent pulses [29] is still challenging task. In the future, understanding and, eventually, controlling the intriguing nature of superfluorescent beatings may permit a rapid quantum operation [35-38] liberated from the rather slow spontaneous emission processes from atoms and molecules.

**Funding.** Robert A. Welch Foundation (A-1547); National University of Mongolia (P2020-3967).

**Disclosures.** The authors declare no conflicts of interest.

**Supplemental document**. See Supplemental Document for supporting content.

# Supplemental Document

Figure S1 (A) shows schematics of an input laser beam enters the Cs atomic vapor and produces a blue light. The Cs atomic energy level diagram is shown in Fig. S1 (B), where the process on transitions 6S - 8S - 7P - 6S is depicted. The emissions in transitions $6S_{1/2}$ – $7P_{3/2}$ at 456 nm and $6S_{1/2}$ – $7P_{1/2}$ at 459 nm are spectrally resolved, see the spectral data in Fig. S1 (C). The generated blue light beam profiles of approximately 4 cm diameter are shown for several different input laser powers (from 100 mW to 450 mW with an increment of 50 mW) at the temperature 242 C. In Fig. S1 (E), the intensity of blue light as a function of input laser power is shown. For higher power, beam profiles have distinct ring and central spot structures.

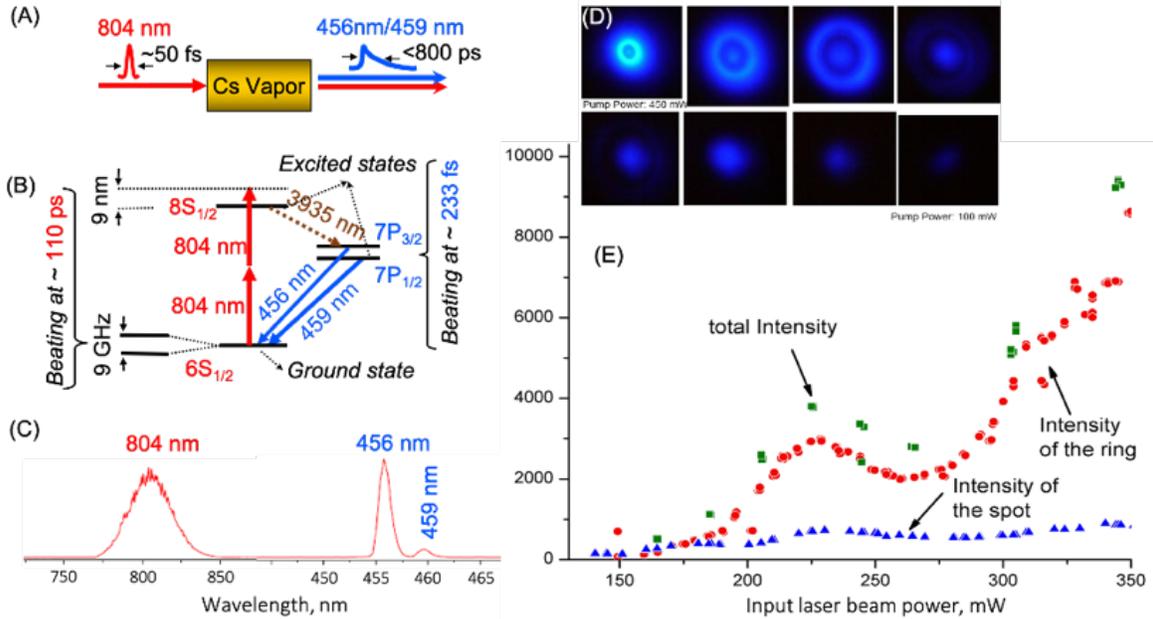

Fig. S1. Supplementary Information. (A) A 60-fs long pulses with 804 nm center wavelength at 1 kHz repetition rate excite cesium atoms and produce a blue light. (B) The transient four-wave mixing process: The two-photon excitation is followed by the simultaneous emissions in the upper (SF) and lower (YSF) transitions. (C). Spectra of the input (804 nm) and output pulses (456/459 nm). (D) The output beam profiles at various input power from 100 mW to 450 mW. (E) The blue light intensity as a function of input power.

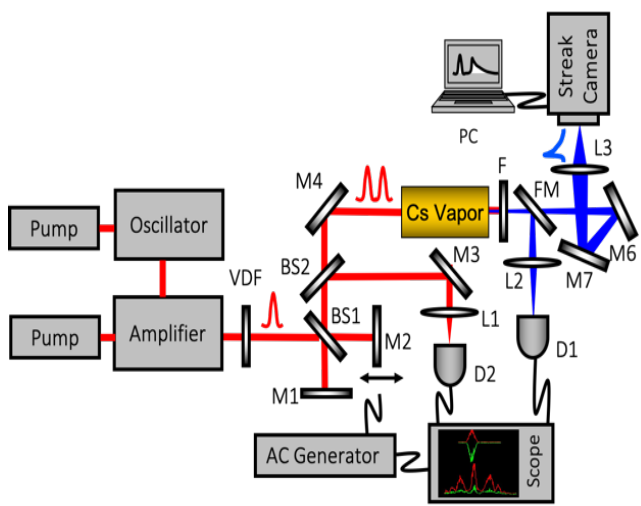

Fig. S2. Supplementary Information. Experimental setup layout. M1-7: mirrors, PC – computer, BS1-2: beamsplitters, D1-2: photodiode detectors, L1-3: lenses, FM: flip mirror, VDF: variable neutral density filter, F: color filters, Cs: cesium, Ti:Sapphire 1 kHz pulse repetition rate laser system including two pump lasers, an oscillator and regenerative amplifier. M2 mirror is attached to a scanner controlled by AC generator. A scope and PC read out the designated output signals.

A setup layout is sketched in Fig. S2, where the amplifier output pulses with an additional autocorrelation measurement option (with the help of the Michelson interferometer) that produces double pulses with a variable delay enter the cell and generate blue light. The generated blue light is spectrally filtered and then detected either by the specially designed photodiode detectors or the 2-ps resolution streak camera. Commercially available photodiode detectors are used (OPT101, Thorlabs Inc.). Its response time is intentionally increased to a half millisecond as to improve signal to noise ratio by the time-integrated detection scheme. A digital storage oscilloscope (Tektronix) displays and records both input and generated light. A fast streak camera (Hamamatsu, c5680) is used to record temporal characteristics of the input and generated light.

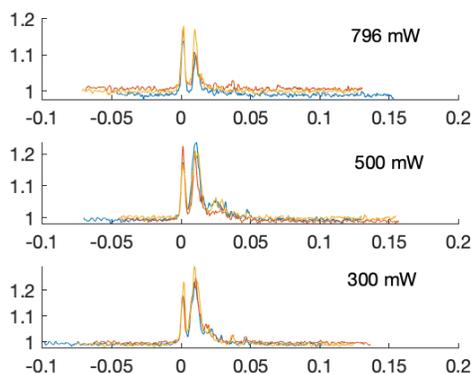

Fig. S3. Supplementary Information. Data taken at temperature 242 C and high input powers for 500 mW and 796 mW fand compared to the data for 300 mW power as functions of time in nanoseconds.

Figure S3 displays data at temperature 242 C. The measured pulse shapes and delay offsets become insensitive for relatively high input powers at least within a 2 ps resolution.